\documentclass{elsart3}

\usepackage{graphicx}

\usepackage{amssymb}

\begin{document}

\begin{frontmatter}



\title{INVESTIGATION OF SCALING PROPERTIES OF HYSTERESIS IN FINEMET THIN FILMS}

\author[UFSM,IEN]{L. Santi},
\author[UFSM]{L. S. Dorneles},
\author[UFSM]{R.L.Sommer},
\author[ROM]{F.Colaiori},
\author[ROM]{S. Zapperi},
\author[IEN]{A. Magni}, and
\author[IEN]{G. Durin\corauthref{cor1}}

\address[UFSM]{Univ. Federal de Santa Maria, Dep. de Fisica, UFSM, 97105-900, Santa Maria, RS, Brasil}
\address[ROM]{INFM Unit\`{a} di Roma 1, Dip. di Fisica, Universit\`{a} "La Sapienza", P.le A. Moro 2,
        00185 Roma, Italy}
\address[IEN]{IEN Galileo Ferraris, str. delle Cacce 91, 10135 Torino, Italy}
\corauth[cor1]{Corresponding author. Tel.: +39 0113919841; fax: +39 0113919834; e-mail:
durin@ien.it}

\begin{abstract}
We study the behavior of hysteresis loops in Finemet
Fe$_{73.5}$Cu$_1$Nb$_3$Si$_{18.5}$B$_4$ thin films by using a fluxometric setup based
on a couple of well compensated pickup coils. The presence of scaling laws of the
hysteresis area is investigated as a function of the amplitude and frequency of the
applied field, considering sample thickness from about 20 nm to 5 $\mu$m. We do not
observed any scaling predicted by theoretical models, while dynamic loops show a
logarithmic dependence on the frequency.
\end{abstract}

\end{frontmatter}


The hysteresis properties of magnetic thin and ultrathin films has been extensively
investigated both theoretically and experimentally \cite{CHA-99,RUI-02,JAN-01}. In view
of their application in magnetic devices, it is fundamental to understand the
microscopical origin of magnetization processes, especially considering the detailed
features of domain wall motion and domain nucleation. Large part of present literature
has analyzed the behavior of dynamic hysteresis measured by using the longitudinal
magneto-optical Kerr effect (MOKE). Following many theoretical studies, data are
analyzed in order to verify the existence of universal scaling laws. In particular, the
dynamic hysteresis loop area $A$ is assumed to follow a law of the type $A \sim A_0 +
H_0^\alpha \omega^\beta$, where $A_0$ is the static hysteresis, $H_0$ and $\omega$ are
the amplitude and the frequency of the applied field, and $\alpha, \beta$ two critical
exponents. As a matter of fact, the estimated critical exponents span a quite large
range, so that the true existence of scaling laws and universality is still under
question. Beside this, a direct connection between experimental observations and the
underlying magnetization processes cannot be still provided.

We investigate the dynamical hysteresis properties from a technique complementary to
the MOKE. Using a couple of well compensated sets of coils, we detect the induced flux
rate, which is proportional to the volume magnetization change. On the contrary, the
MOKE technique detects the magnetization only on the sample surface and within the
laser spot area. We believe that the use of both techniques helps to better investigate
the magnetization processes, especially when the sample thickness can play an important
role.
\begin{figure}
\centering \includegraphics[width=7cm]{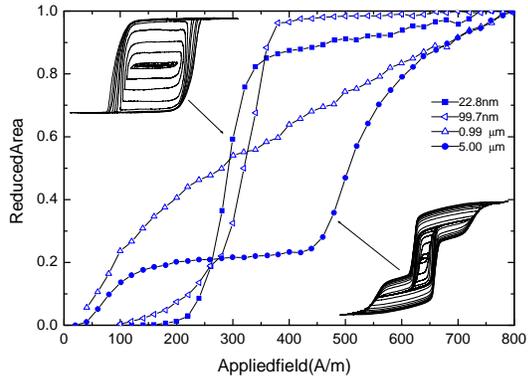}

\caption{Hysteresis loop area for Finemet thin films at different thicknesses. The area
is divided by the value at H=800 A/m for convenience. Examples of minor loop
measurements are shown for the thinnest (22.8 nm) and the thickest (5 $\mu$m) sample. }
\label{fig:A_vs_H}
\end{figure}


We consider a set of amorphous Finemet thin films of composition
Fe$_{73.5}$Cu$_1$Nb$_3$Si$_{18.5}$B$_4$, with thickness ranging from about 20 nm to 5
$\mu$m, calibrated with low angle x-ray diffractometry. Lateral dimensions are about
1.5 cm by 3 cm. The dependence of the hysteresis loop on the applied field amplitude is
reported in Fig.~\ref{fig:A_vs_H} at four sample thickness. The data show very
different behaviors, and simple power laws does not hold. At lower thickness, the loops
are roughly squared, as shown in the upper left corner of Fig.~\ref{fig:A_vs_H}, and
the loop area steeply changes for applied field approaching the coercive field. Only at
intermediate thickness, we can fit the data with a power law, getting $\alpha \sim
0.4$. Finally, at larger thickness, the loops display a "double switching" structure
(lower right corner), similar to whose reported for some spin-valve structures
\cite{LEE-00}: this behavior suggests the presence of two magnetic sub-systems having
different coercive fields. We do not know the origin of this hysteresis, but we can
confirm this is a common feature of Finemet samples having the same thickness, as we
found similar loops for samples of slightly different composition.

\begin{figure}
\centering \includegraphics[width=7cm]{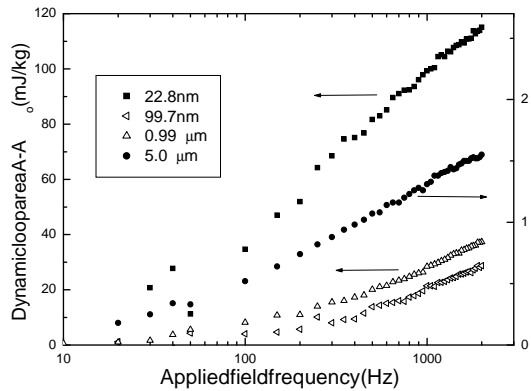}

\caption{Frequency dependence of the dynamic hysteresis area $A - A_0$ obtained by
fluxometric measurements. Data for the thickest sample are obtained considering only
the inner loops not showing the double switching as in Fig.~\ref{fig:A_vs_H}.}
\label{fig:A_vs_fr}
\end{figure}

A similar dependence on the sample thickness is not found when considering the behavior
of dynamic loops: in this case, all the hysteresis area roughly follow the logarithm of
frequency, as shown in the semi-log plot of Fig.~\ref{fig:A_vs_fr}. It is interesting
to observe the results for the 5 $\mu$m; the double switching loops are not much
affected by the frequency, as we observe only a slight change in the curves around the
coercive field. This is in contrast with the results in the spin-valves of
Ref.~\cite{LEE-00} where the loops transform to a "single switch" behavior. We thus
considered only the inner loops not showing the double switching, and found they also
show the logarithmic dependence as the other samples, as shown in
Fig.~\ref{fig:A_vs_fr}. Even if the data at low frequencies are rather noisy, as usual
with this fluxometric technique, we cannot observe any clear dynamic transition in the
hysteresis, which would imply a steep change of exponent $\beta$. All these results
suggest that a more general approach, not only based on the existence of scaling laws,
is needed to investigate the hysteresis properties in thin films.

\end{document}